\documentstyle[prd,aps,preprint,tighten,epsfig]{revtex}

\begin{document}

\draft

\title{Vanishing Effective Mass of the Neutrinoless Double Beta Decay?}
\author{{\bf Zhi-zhong Xing}}
\address{CCAST (World Laboratory), P.O. Box 8730, Beijing 100080, China \\
and Institute of High Energy Physics, Chinese Academy of Sciences, \\
P.O. Box 918 (4), Beijing 100039, China 
\footnote{Mailing address} \\
({\it Electronic address: xingzz@mail.ihep.ac.cn}) } 
\maketitle

\begin{abstract}
We stress the point that massive neutrinos may be Majorana particles 
even if the effective mass of the neutrinoless double beta decay 
$\langle m\rangle_{ee}$ vanishes. We show that current neutrino
oscillation data {\it do} allow $\langle m\rangle_{ee} = 0$ to hold, 
if the Majorana CP-violating phases lie in two specific regions. 
Strong constraints on the neutrino mass spectrum can then be obtained. 
A possible texture of the neutrino mass matrix is also illustrated 
under the $\langle m\rangle_{ee} = 0$ condition.
\end{abstract}


\newpage

The recent SK \cite{SK}, SNO \cite{SNO}, 
KamLAND \cite{KM} and K2K \cite{K2K} experiments have 
provided us with very convincing evidence that the solar and
atmospheric neutrino anomalies are both due to neutrino
oscillations. The occurence of neutrino oscillations implies
that neutrinos are massive and lepton flavors are mixed. If
neutrinos are Majorana particles, a complete parametrization
of the $3\times 3$ lepton flavor mixing matrix requires 
three mixing angles, one Dirac-type CP-violating phase and
two Majorana-type CP-violating phases \cite{FX01}. While
three mixing angles have been determined or constrained by 
current neutrino oscillation data to an acceptable degree
of accuracy, three CP-violating phases are entirely 
unrestricted. It is expected that the Dirac phase can be measured 
from CP- or T-violating effects in the long-baseline neutrino 
oscillation experiments \cite{LBL}. To measure two Majorana 
phases is extremely difficult, because all possible
lepton-number-nonconserving processes induced by light Majorana
neutrinos are strongly suppressed in magnitude \cite{Kayser}.

The most sensitive way to get some information on
two Majorana CP-violating phases is to detect the neutrinoless 
double beta decay of some even-even nuclei,
\begin{equation}
A(Z,N) \; \rightarrow \; A(Z+2, N-2) + 2e^- \; ,
%       (1)
\end{equation}
which can occur through the exchange of a Majorana neutrino between 
two decaying neutrons inside a nucleus, as illustrated in Fig. 1.
It would be forbidden, however, if neutrinos were Dirac particles. 
Thus the neutrinoless double beta decay provides us with a unique 
opportunity to identify the Majorana nature of massive neutrinos.
The rate of the neutrinoless double beta decay is proportional to
an effective neutrino mass term, defined as 
\begin{equation}
\langle m \rangle_{ee} \; = \; \left |
m_1 V^2_{e1} + m_2 V^2_{e2} + m_3 V^2_{e3} \right | \; ,
%       (2)
\end{equation}
where $m_i$ (for $i=1,2,3$) denote the physical masses of three
neutrinos, and $V_{ei}$ stand for the elements in the first row
of the $3\times 3$ lepton flavor mixing matrix $V$. It is obvious
that $\langle m\rangle_{ee} = 0$ would trivially hold, if $m_i = 0$ 
were taken.

While $\langle m\rangle_{ee} \neq 0$ must imply that neutrinos
are Majorana particles, $\langle m\rangle_{ee} = 0$ does not 
{\it necessarily} imply that neutrinos are Dirac particles. The
reason is simply that the Majorana phases hidden in $V_{ei}$ may
lead to significant cancellations on the right-hand side of 
Eq. (2), making $\langle m\rangle_{ee}$ vanishing or too small to
be detectable \cite{B}. Hence much care has to be taken, if no 
convincing signal of the neutrinoless double beta decay can be 
experimentally established
%%%%%%%%%%%%%%%%%%%%%%%%%%%
\footnote{Klapdor-Kleingrothaus {\it et al.} have recently reported
the first evidence for the neutrinoless double beta decay \cite{KK}.
However, their result was criticized by some authors \cite{KK2}.
Future experiments will have sufficiently high sensitivity to 
clarify the present debates \cite{KK3}, to confirm or to 
disprove the alleged result in Ref. \cite{KK}.}:
%%%%%%%%%%%%%%%%%%%%%%%%%%%
it may imply that (1) the experimental sensitivity is not high
enough; (2) the massive neutrinos are Dirac particles; or (3) the
vanishing or suppression of $\langle m\rangle_{ee}$ is due to
large cancellations induced by the Majorana CP-violating phases.
The third possibility is certainly interesting and 
important \cite{Vissani}, and it deserves to be carefully examined 
from a model-independent point of view and with the help of the
latest experimental data.

The main purpose of this paper is to find out the parameter
space of two Majorana phases in the case of 
$\langle m\rangle_{ee} = 0$. We demonstrate that current 
neutrino oscillation data {\it do} allow 
$\langle m\rangle_{ee} = 0$ to hold, if the Majorana phases lie
in two specific regions. Very strong constraints on three neutrino
masses can then be obtained. We find that the neutrino mass 
spectrum performs a normal hierarchy: $m_1 < m_2 < m_3$. Finally,
we present some brief discussions about how to recast the texture
of the neutrino mass matrix under the $\langle m\rangle_{ee} = 0$ 
condition.

It is clear in Eq. (2) that only the flavor mixing matrix elements
$V_{e1}$, $V_{e2}$ and $V_{e3}$ are relevant to the effective mass
of the neutrinoless double beta decay. Without loss of generality,
one may redefine the phases of three charged lepton fields in an
appropriate way such that the phases of $V_{e1}$ and $V_{e2}$ are
purely of the Majorana type and $V_{e3}$ is real \cite{Xing02}.
In other words,
\begin{equation}
\arg (V_{e1}) = \rho \; , ~~~~
\arg (V_{e2}) = \sigma \; , ~~~~
\arg (V_{e3}) = 0 \; . 
%       (3)
\end{equation}
Note that $\rho$ and $\sigma$ have nothing to do with CP and T
violation in normal neutrino oscillations. Taking account of
Eqs. (2) and (3), we find that 
$\langle m\rangle_{ee} = 0$ requires
\begin{eqnarray}
&& m_1 |V_{e1}|^2 \sin 2\rho + m_2 |V_{e2}|^2 \sin 2\sigma
\; =\; 0 \; , ~~~~~~
\nonumber \\
~~~~~~~~ && m_1 |V_{e1}|^2 \cos 2\rho + m_2 |V_{e2}|^2 \cos 2\sigma
+ m_3 |V_{e3}|^2 \; =\; 0 \; . ~~~~~~
%       (4)
\end{eqnarray}
These two conditions, together with current experimental data
on the flavor mixing matrix elements ($|V_{e1}|$, $|V_{e2}|$
and $|V_{e3}|$) and the mass-squared differences of solar and
atmospheric neutrino oscillations 
($\Delta m^2_{\rm sun} \equiv |m^2_2 - m^2_1|$ and
$\Delta m^2_{\rm atm} \equiv |m^2_3 - m^2_2|$), allow us to
determine or constrain both the masses of three neutrinos 
($m_1$, $m_2$ and $m_3$) and the Majorana phases of CP violation
($\rho$ and $\sigma$). More specific discussions about the
consequences of Eq. (4) are in order.

(a) Note that the present solar neutrino data support 
$0\leq m_1 < m_2$ \cite{SNO,KM}. If $m_1 =0$ holds, then Eq. (4) 
requires $\sigma = (2n+1)\pi/2$ with $n=0,1,2$ $\cdot\cdot\cdot$ for
arbitrary values of $\rho$. In this case, we immediately obtain
\begin{equation}
\frac{m_2}{m_3} \; =\; \frac{|V_{e3}|^2}{|V_{e2}|^2} \;\; , ~~~~~~~
\frac{\Delta m^2_{\rm sun}}{\Delta m^2_{\rm atm}} \; =\;
\frac{|V_{e3}|^4}{|V_{e2}|^4 - |V_{e3}|^4} \;\; .
%       (5)
\end{equation}
Because of $|V_{e3}| < |V_{e2}|$, the neutrino mass spectrum 
performs a normal hierarchy (i.e., $m_3 > m_2 > m_1 =0$). The
absolute values of $m_2$ and $m_3$ are given by
\begin{equation}
~~ m_2 = \frac{|V_{e3}|^2 \sqrt{\Delta m^2_{\rm atm}}}
{\sqrt{|V_{e2}|^4 - |V_{e3}|^4}} \;\; , ~~~
m_3 = \frac{|V_{e2}|^2 \sqrt{\Delta m^2_{\rm atm}}}
{\sqrt{|V_{e2}|^4 - |V_{e3}|^4}} \;\; . ~~
%       (6)
\end{equation}
Our numerical analysis will show that the possibility 
$m_1 = \langle m\rangle_{ee} =0$ is actually allowed by current
neutrino oscillation data.

(b) Whether $0< m_2 < m_3$ or $0 \leq m_3 < m_2$ holds remains 
an open question. If $m_3 = 0$ held, then Eq. (4) would require
$|\rho - \sigma| = (2n+1)\pi/2$ with $n$ being an arbitrary
integer. In this case, we would be led to
\begin{equation}
\frac{m_1}{m_2} \; =\; \frac{|V_{e2}|^2}{|V_{e1}|^2} \;\; , ~~~~~~~
\frac{\Delta m^2_{\rm sun}}{\Delta m^2_{\rm atm}} \; =\;
\frac{|V_{e1}|^4 - |V_{e2}|^4}{|V_{e1}|^4} \;\; ;
%       (7)
\end{equation}
as well as 
\begin{equation}
~~ m_1 = \frac{|V_{e2}|^2 \sqrt{\Delta m^2_{\rm sun}}}
{\sqrt{|V_{e1}|^4 - |V_{e2}|^4}} \;\; , ~~~
m_2 = \frac{|V_{e1}|^2 \sqrt{\Delta m^2_{\rm sun}}}
{\sqrt{|V_{e1}|^4 - |V_{e2}|^4}} \;\; . ~~
%       (8)
\end{equation}
Our numerical analysis will show that the possibility
$m_3 = \langle m\rangle_{ee} = 0$ has definitely been ruled 
out by current neutrino oscillation data.

(c) If $m_1 \neq 0$ but $\sin 2\rho =0$ or $\sin 2\sigma = 0$ 
holds, then Eq. (4) requires $\rho = n\pi$ and
$\sigma = (2n'+1)\pi/2$ with
\begin{equation}
m_1 |V_{e1}|^2 + m_3 |V_{e3}|^2 \; =\; m_2 |V_{e2}|^2 \; ;
%       (9)
\end{equation}
or $\rho = (2n+1)\pi/2$ and $\sigma = n'\pi$ with
\begin{equation}
m_2 |V_{e2}|^2 + m_3 |V_{e3}|^2 \; =\; m_1 |V_{e1}|^2 \; ;
%       (10)
\end{equation}
or $\rho = (2n+1)\pi/2$ and $\sigma = (2n'+1)\pi/2$ with
\begin{equation}
m_1 |V_{e1}|^2 + m_2 |V_{e2}|^2 \; =\; m_3 |V_{e3}|^2 \; ,
%       (11)
\end{equation}
where $n$ and $n'$ are arbitrary integers. With the help of
\begin{eqnarray}
m_1 & = & \sqrt{m^2_3 \pm \Delta m^2_{\rm atm} + 
\Delta m^2_{\rm sun}} \;\; , ~~~~
\nonumber \\
m_2 & = & \sqrt{m^2_3 \pm \Delta m^2_{\rm atm}} \;\; , 
%       (12)
\end{eqnarray}
$m_3$ can be solved from Eq. (9), (10) or (11). Thus the
spectrum of three neutrino masses is fully determinable. 
Our numerical analysis will show that Eqs. (9) and (10) are
consistent quite well with current neutrino oscillation data.
In comparison, Eq. (11) is also allowed but it is less favored.  

(d) Besides the special cases considered above, more general
results for $(m_1, m_2, m_3)$ and $(\rho, \sigma)$ can be 
obtained from Eq. (4). Indeed,
\begin{eqnarray}
\frac{m_1}{m_2} & = & -\frac{|V_{e2}|^2}{|V_{e1}|^2}
\cdot \frac{\sin 2\sigma}{\sin 2\rho} \;\; ,
\nonumber \\
\frac{m_2}{m_3} & = & +\frac{|V_{e3}|^2}{|V_{e2}|^2}
\cdot \frac{\sin 2\rho}{\sin 2 (\sigma - \rho)} \;\; .
%       (13)
\end{eqnarray}
Since $0 \leq m_1/m_2 <1$ and $0 < m_2/m_3$ hold, part of the
$(\rho, \sigma)$ parameter space must be excluded. Furthermore,
we arrive at
\begin{equation}
\frac{\Delta m^2_{\rm sun}}{\Delta m^2_{\rm atm}}
\; =\; \frac{|V_{e3}|^4}{|V_{e1}|^4} \cdot
\frac{\displaystyle \left | |V_{e1}|^4 \sin^2 2\rho -
|V_{e2}|^4 \sin^2 2\sigma \right |}
{\displaystyle \left | |V_{e2}|^4 \sin^2 2(\sigma -\rho) -
|V_{e3}|^4 \sin^2 2\rho \right |} \;\; .
%      (14)
\end{equation}
Note that the lepton flavor mixing matrix elements $|V_{e1}|^2$, 
$|V_{e2}|^2$, $|V_{e3}|^2$, $|V_{\mu 3}|^2$ and $|V_{\tau 3}|^2$ 
are associated respectively with the mixing factors of solar, 
atmospheric and CHOOZ \cite{CHOOZ} reactor neutrino oscillations,
\begin{eqnarray}
\sin^2 2\theta_{\rm sun} & = & 4 |V_{e1}|^2 |V_{e2}|^2 \; ,
\nonumber \\
\sin^2 2\theta_{\rm atm} & = & 4 |V_{\mu 3}|^2 
\left (1 - |V_{\mu 3}|^2 \right ) \; ,
\nonumber \\
\sin^2 2\theta_{\rm chz} & = & 4 |V_{e3}|^2 
\left (1 - |V_{e3}|^2 \right ) \; .
%       (15)
\end{eqnarray}
Reversely, we have \cite{Xing03}
\begin{eqnarray}
|V_{e1}|^2 & = & \frac{1}{2} \left ( \cos^2\theta_{\rm chz}
+ \sqrt{\cos^4\theta_{\rm chz} - \sin^2 2\theta_{\rm sun}} \right ) \; ,
\nonumber \\
|V_{e2}|^2 & = & \frac{1}{2} \left ( \cos^2\theta_{\rm chz}
- \sqrt{\cos^4\theta_{\rm chz} - \sin^2 2\theta_{\rm sun}} \right ) \; ,
\nonumber \\
|V_{e3}|^2 & = & \sin^2 \theta_{\rm chz} \; ,
\nonumber \\
|V_{\mu 3}|^2 & = & \sin^2 \theta_{\rm atm} \; ,
\nonumber \\
|V_{\tau 3}|^2 & = & \cos^2 \theta_{\rm chz} - 
\sin^2 \theta_{\rm atm} \; .
%	(16)
\end{eqnarray}
In view of the recent SK \cite{SK}, SNO \cite{SNO}, 
KamLAND \cite{KM}, K2K \cite{K2K} and CHOOZ \cite{CHOOZ}
data on neutrino oscillations, we have 
$\Delta m^2_{\rm sun} \in [5.9, ~ 8.8] \times 10^{-5} ~ {\rm eV}^2$,
$\sin^2 \theta_{\rm sun} \in [0.25, ~ 0.40]$ \cite{Fit};
$\Delta m^2_{\rm atm} \in [1.65, ~ 3.25] \times 10^{-3} ~ {\rm eV}^2$,
$\sin^2 2\theta_{\rm atm} \in [0.88, ~ 1.00]$ \cite{Fogli}; and
$\sin^2 2\theta_{\rm chz} < 0.2$ at the $90\%$ confidence level.
With the help of these experimental results, the allowed ranges 
of $\rho$ and $\sigma$ can then be obtained from Eqs. (14) and (16).

We plot the parameter space of two Majorana phases in Fig. 2(a).
It becomes obvious that $\rho$ and $\sigma$ may take many nontrivial
values, which guarantee $\langle m\rangle_{ee} =0$. This important
point has not been observed before. Note that the $(\rho, \sigma)$
parameter space in Fig. 2(a) can be generalized to the larger
$(\rho \pm n_1\pi, \sigma \pm n_2\pi)$ parameter space, where $n_1$ 
and $n_2$ are arbitrary integers. For illustration, we typically pick 
\begin{equation}
(\rho, ~\sigma) \; = \; \left (\frac{\pi}{4}, ~\frac{2\pi}{3} \right)
\;\;\;\;\; {\rm or} \;\;\;\;\; \left (\frac{3\pi}{4}, ~\frac{\pi}{3} \right)
%       (17)
\end{equation}
from Fig. 2(a). Then we arrive at
\begin{equation}
\frac{m_1}{m_2} \; = \; 
\frac{\sqrt{3}}{2} \frac{|V_{e2}|^2}{|V_{e1}|^2} \;\; ,
~~~~~
\frac{m_2}{m_3} \; = \; 2 \frac{|V_{e3}|^2}{|V_{e2}|^2} \;\; .
%       (18)
\end{equation}
Given the ranges of $\theta_{\rm sun}$ and $\theta_{\rm chz}$ favored 
by current experimental data, $m_1 < m_2$ and $m_2 < m_3$ are found to
hold from Eq. (18). Thus three neutrino masses perform a normal 
hierarchy in this specific but interesting case.
 
A detailed analysis of the $(m_1/m_2, m_2/m_3)$ parameter space is
shown in Fig. 2(b). We find that two specific regions of 
$\rho$ and $\sigma$ in Fig. 2(a) correspond to a common region
of $m_1/m_2$ and $m_2/m_3$ in Fig. 2(b), just like the specific 
case illustrated in Eqs. (17) and (18). One can see that both
$0\leq m_1/m_2 <1$ and $0< m_2/m_3 <1$ hold, 
thus three neutrino masses have a normal hierarchy 
(i.e., $m_1 < m_2 < m_3$). When $m_1/m_2$ approaches 1,
$m_2/m_3$ is somehow close to 1 too. In this case, which is not
very likely, three neutrino masses are nearly degenerate 
(i.e., $m_1 \approx m_2 \approx m_3$). Note that $m_2/m_3$ and
$|V_{e3}|$ have minimal values $(m_2/m_3)_{\rm min} \approx 0.135$ 
and $(|V_{e3}|)_{\rm min} \approx 0.0695$, respectively. The
latter, which corresponds to 
$(\theta_{\rm chz})_{\rm min} \approx 4^\circ$
or $(\sin^2 2\theta_{\rm chz})_{\rm min} \approx 0.02$, 
is associated with the prerequisite $\langle m\rangle_{ee} =0$. 
In contrast, we do not find any restriction on the input and 
output values of $\theta_{\rm sun}$ or $\theta_{\rm atm}$ in 
our numerical calculation.

We see that $m_2/m_3$ is most likely to be around 0.2, implying 
that $m_3 \approx \sqrt{\Delta m^2_{\rm atm}} \sim 0.05 ~ {\rm eV}$ 
should be a good approximation. In this case, the effective mass
of the tritium beta decay $\langle m\rangle_e$ is too small to
be detected by the KATRIN experiment \cite{No}. If 
$m_2/m_3 \sim 0.9$ is taken, one will get $m_3 \sim 0.1$ eV and 
$m_1 + m_2 + m_3 \sim 0.3$ eV, consistent with the recent WMAP data
$m_1 + m_2 + m_3 < 0.71$ eV \cite{WMAP}.  

Let us proceed to discuss how to recast the neutrino mass matrix 
under the condition $\langle m\rangle_{ee} =0$. In the flavor basis 
where the charged lepton mass matrix is diagonal, the neutrino
mass matrix $M$ can be written as
\begin{equation}
M \; =\; V \left ( \matrix{
m_1 & 0 & 0 \cr
0 & m_2 & 0 \cr
0 & 0 & m_3 \cr} \right ) V^T \; .
%       (19)
\end{equation}
The lepton flavor mixing matrix $V$ consists of three nontrivial
CP-violating phases: the Dirac phase $\delta$ and the Majorana
phases $\rho$ and $\sigma$. For simplicity, we assume $\delta =0$
to examine the dependence of $M$ on $\rho$ and $\sigma$. Then
$V$ may take the form
\begin{equation}
V \; =\; \left ( \matrix{
|V_{e1}| & |V_{e2}| & |V_{e3}| \cr
-|V_{\mu 1}| & |V_{\mu 2}| & |V_{\mu 3}| \cr
|V_{\tau 1}| & -|V_{\tau 2}| & |V_{\tau 3}| \cr}
\right ) 
\left ( \matrix{
e^{i\rho} & 0 & 0 \cr
0 & e^{i\sigma} & 0 \cr
0 & 0 & 1 \cr} \right ) \; ,
%       (20)
\end{equation}
in which $|V_{e1}|$, $|V_{e2}|$, $|V_{e3}|$, $|V_{\mu 3}|$ and
$|V_{\tau 3}|$ have been given in Eq. (16) in terms of 
$\theta_{\rm sun}$, $\theta_{\rm atm}$ and $\theta_{\rm chz}$;
and $|V_{\mu 1}|$, $|V_{\mu 2}|$, $|V_{\tau 1}|$ and $|V_{\tau 2}|$
read as follows (in the assumption of $\delta =0$ \cite{Xing03}):
\begin{eqnarray}
|V_{\mu 1}| & = & \frac{|V_{e2}| |V_{\tau 3}| + |V_{e1}|
|V_{e3}| |V_{\mu 3}|}{1 - |V_{e3}|^2} \;\; ,
\nonumber \\
|V_{\mu 2}| & = & \frac{|V_{e1}| |V_{\tau 3}| - |V_{e2}|
|V_{e3}| |V_{\mu 3}|}{1 - |V_{e3}|^2} \;\; ,
\nonumber \\
|V_{\tau 1}| & = & \frac{|V_{e2}| |V_{\mu 3}| - |V_{e1}|
|V_{e3}| |V_{\tau 3}|}{1 - |V_{e3}|^2} \;\; ,
\nonumber \\
|V_{\tau 2}| & = & \frac{|V_{e1}| |V_{\mu 3}| + |V_{e2}|
|V_{e3}| |V_{\tau 3}|}{1 - |V_{e3}|^2} \;\; .
%       (21)
\end{eqnarray}
With the help of Eqs. (19) and (21), six independent elements of 
the symmetric neutrino mass matrix $M$ can be expressed as
\begin{equation}
\left ( \matrix{
M_{ee} \cr M_{e \mu} \cr M_{e \tau} \cr M_{\mu \mu} \cr
M_{\mu \tau} \cr M_{\tau \tau} \cr} \right ) \; =\;
m_1 \left ( \matrix{
|V_{e1}|^2 \cr -|V_{e1}| |V_{\mu 1}| \cr |V_{e1}| |V_{\tau 1}| \cr
|V_{\mu 1}|^2 \cr -|V_{\mu 1}| |V_{\tau 1}| \cr |V_{\tau 1}|^2 \cr}
\right ) e^{2i\rho} + m_2 \left ( \matrix{
|V_{e2}|^2 \cr |V_{e2}| |V_{\mu 2}| \cr -|V_{e2}| |V_{\tau 2}| \cr
|V_{\mu 2}|^2 \cr -|V_{\mu 2}| |V_{\tau 2}| \cr |V_{\tau 2}|^2 \cr}
\right ) e^{2i\sigma} + m_3 \left ( \matrix{
|V_{e3}|^2 \cr |V_{e3}| |V_{\mu 3}| \cr |V_{e3}| |V_{\tau 3}| \cr
|V_{\mu 3}|^2 \cr |V_{\mu 3}| |V_{\tau 3}| \cr |V_{\tau 3}|^2 \cr}
\right ) \; .
%       (22)
\end{equation}
As $|M_{ee}| = \langle m\rangle_{ee}$ holds, we have $M_{ee} =0$
under the condition $\langle m\rangle_{ee} =0$. For illustration,
we calculate the other five matrix elements in Eq. (22) by using
the typical values of $\rho$ and $\sigma$ chosen in Eq. (17) and
taking $\theta_{\rm sun} = 33^\circ$, $\theta_{\rm atm} = 45^\circ$ 
and $\theta_{\rm chz} = 9^\circ$. The numerical result are
$m_1/m_2 \approx 0.42$, $m_2/m_3 \approx 0.15$, and
\begin{equation}
M \; \approx \; m_3 \left ( \matrix{
{\bf 0} & 0.11 e^{\mp i 36^\circ} & 0.15 e^{\pm i 26^\circ} \cr
0.11 e^{\mp i 36^\circ} & 0.48 e^{\mp i 2.3^\circ} &
0.51 e^{\pm i 3.8^\circ} \cr
0.15 e^{\pm i 26^\circ} & 0.51 e^{\pm i 3.8^\circ} &
0.45 e^{\mp i 6.5^\circ} \cr} \right ) \; ,
%       (23)
\end{equation}
where $m_3 \sim 0.05$ eV. We see that this one-zero texture of 
$M$ does not perform an apparent hierarchy and has little
similarity with those two-zero textures of $M$ illustrated in
Ref. \cite{Xing02a}. Of course, the form of $M$ in Eq. (23) will
get modified, if the Dirac CP-violating phase of $V$ is switched
on. A more general and delicate analysis of possible patterns of 
$M$ under the $\langle m\rangle_{ee} =0$ condition will be done
elsewhere. 

It is worth mentioning that the $\langle m\rangle_{ee} =0$
condition may be taken as a prerequisite to build models
for the Majorana neutrino mass matrix $M$, if the neutrinoless 
double beta decay is unable to be detected. Another two empirical
conditions, ${\rm Det}M =0$ \cite{Yanagida} and 
$\lambda_1 + \lambda_2 + \lambda_3 =0$ (where
$|\lambda_i| = m_i$ \cite{He}), have also been discussed to
constrain the form of $M$ in a phenomenological way. Such 
``zero'' conditions are similar to those taken for the matrix
elements of $M$ \cite{Xing02a,F}, in order to reduce the number of
free parameters in $M$. This is one of a few realistic 
approaches \cite{Review}, towards some deeper understanding of 
lepton mass generation and flavor mixing.

In summary, we have stressed the point that 
$\langle m\rangle_{ee} =0$ does not necessarily indicate the
Dirac nature of light neutrinos. Current neutrino oscillation
data do allow $\langle m\rangle_{ee} =0$ to hold, if neutrinos
are Majorana particles and their two Majorana CP-violating 
phases lie in two specific regions. This observation will be
important for model building, in order to completely understand 
or interpret current and future experimental data on the neutrino 
mass spectrum and the neutrinoless double beta decay.

\vspace{0.5cm}

I am grateful to C. Giunti and F. Vissani for useful comments 
and discussions, and to F. Vissani and W. Rodejohann for bringing 
a few relevant references to my attention. I am also indebted to
W.L. Guo for his help in dealing with the figures. This work was 
supported in part by the National Natural Science Foundation of China.

\newpage

%%%%%%%%%%%%%%%%%%%% Fig. 1 %%%%%%%%%%%%%%%%
\begin{figure}[t]
\vspace{-1.2cm}
\epsfig{file=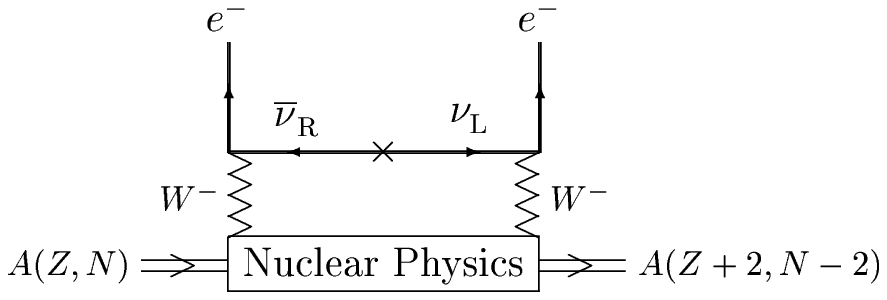,bbllx=3.5cm,bblly=15cm,bburx=18cm,bbury=32.5cm,%
width=15.5cm,height=22cm,angle=0,clip=}
\vspace{-9.8cm}
\caption{Illustrative plot for the neutrinoless double beta decay
of some even-even nuclei via the exchange of a virtual Majorana 
neutrino.}
\end{figure}
%%%%%%%%%%%%%%%%%%%%%%%%%%%%%%%%%%%%%%%%%%%%

\newpage

%%%%%%%%%%%%%%%%%%%% Fig. 2 %%%%%%%%%%%%%%%%
\begin{figure}[t]
\vspace{-2cm}
\epsfig{file=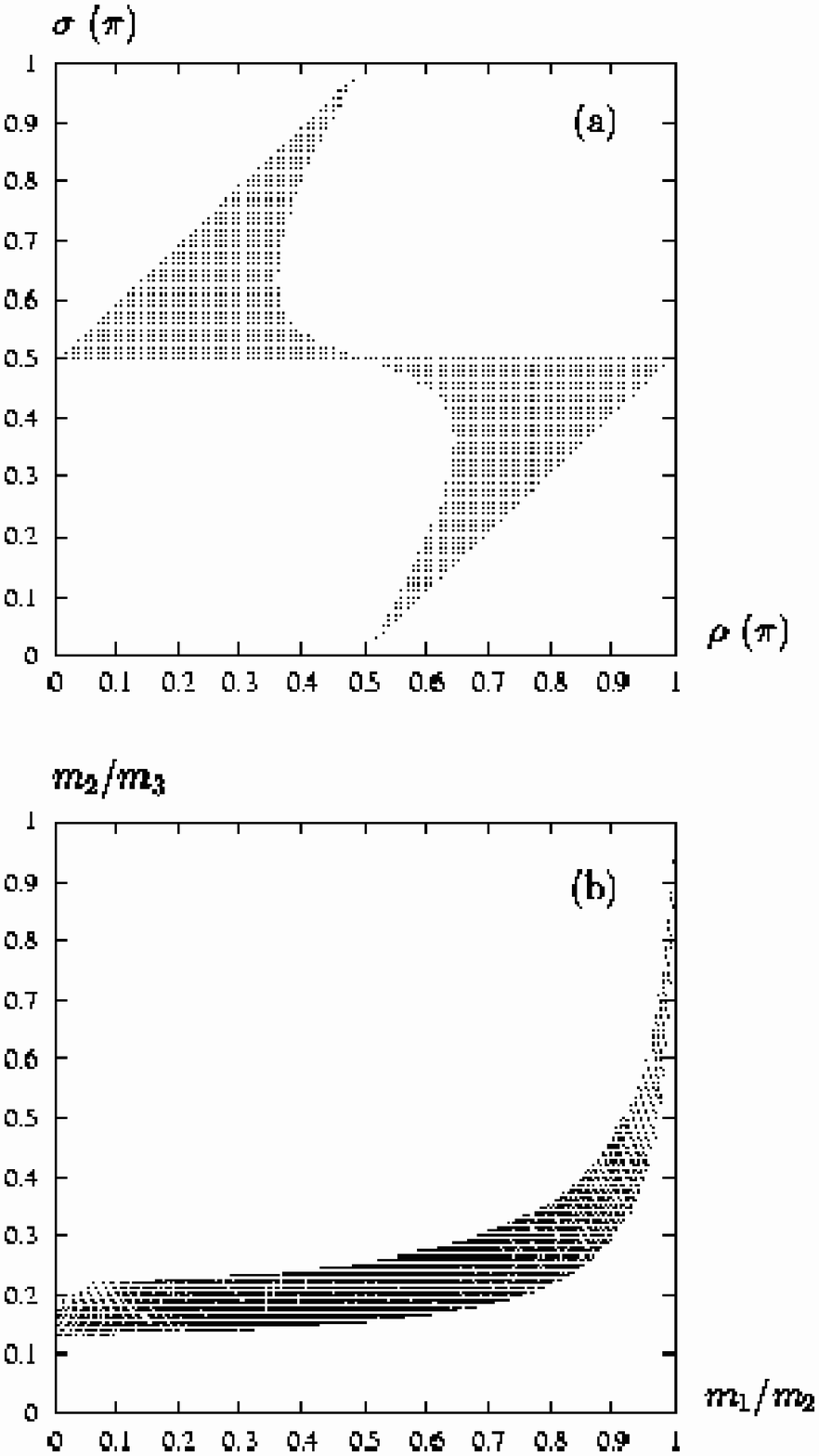,bbllx=-5cm,bblly=-2cm,bburx=18cm,bbury=26cm,%
width=15.5cm,height=20cm,angle=0,clip=}
\vspace{-0.8cm}
\caption{Implications of $\langle m\rangle_{ee} =0$:
the $(\rho, \sigma)$ and $(m_1/m_2, m_2/m_3)$ regions allowed 
by current neutrino oscillation data.}
\end{figure}
%%%%%%%%%%%%%%%%%%%%%%%%%%%%%%%%%%%%%%%%%%%%

\end{document}